\documentclass[12pt]{article}
\input epsf.tex
\newcommand{\beq}{\begin{equation}}
\newcommand{\eeq}{\end{equation}}
\newcommand{\beqa}{\begin{eqnarray}}
\newcommand{\eeqa}{\end{eqnarray}}
\newcommand{\beqar}{\begin{eqnarray*}}
\newcommand{\eeqar}{\end{eqnarray*}}

\newcommand{\eps}{\epsilon}
\newcommand{\ga}{\gamma}

\newcommand{\inn}{\!\cdot\!}

\newcommand{\la}{\lambda}

\newcommand{\z}{\zeta}

\newcommand{\eg}{{\it e.g.,}\ }
\newcommand{\ie}{{\it i.e.,}\ }
\newcommand{\labell}[1]{\label{#1}} %{\label{#1}} %
\newcommand{\reef}[1]{(\ref{#1})}

\newcommand\veps{\varepsilon}

\newcommand\cR{{\cal R}}

\newcommand\tG{{\tilde G}}
\newcommand\tB{{\tilde B}}
\newcommand\tC{{\tilde C}}

\newcommand\tphi{{\tilde \phi}}

\newcommand\ta{{\tilde a}}

\newcommand\ti{{\tilde i}}
\newcommand\tj{{\tilde j}}

\newcommand\Tr{{\rm Tr}}
\newcommand\tr{{\rm tr}}
\begin{document}

\begin{titlepage}

\begin{center}

%\fbox{DRAFT: \today}

%{} \hfill     XX-X-X, \, MIFP-XX-XX

\vskip 2 cm
{\LARGE \bf  Ramond-Ramond field strength \\couplings on D-branes }\\
 
\vskip 1.25 cm
%Katrin Becker\footnote{kbecker@physics.tamu.edu}, 
 Mohammad R. Garousi\footnote{garousi@mail.ipm.ir}  \\
 \vskip 1cm
%{$^1${Department of Physics, Texas A\&M University,\\}{College Station, TX 77843, USA}\\}
\vskip 1 cm
{{\it Department of Physics, Ferdowsi University of Mashhad\\}{\it P.O. Box 1436, Mashhad, Iran}\\}
\vskip .1 cm
\vskip .1 cm
%{{\it School of Physics, Institute for Research in Fundamental Sciences (IPM)\\}{\it P.O. Box 19395-5531, Tehran, Iran}\\}

\end{center}

\vskip 0.5 cm

\begin{abstract}
\baselineskip=18pt
By examining   in details   the disk level S-matrix element of one massless RR and one NSNS states at order $O(\alpha'^2)$, we find the coupling of one RR and one NSNS fields on the world volume of a D$_p$-brane. The  non-zero couplings involve the first derivative of the RR field strengths $F^{(p)}\,, F^{(p+2)}$ and $F^{(p+4)}$. 
We then fix the on-shell ambiguity of the couplings by requiring consistency  with the linear T-duality transformations. Moreover, consistency with the  non-linear T-duality requires that   the RR field strength in the above couplings to be ${\cal F}=d{\cal C}$  where ${\cal C}=e^{B}C$.

\end{abstract}
\vskip 3 cm
\begin{center}
%{\it Dedicated to Farhad Ardalan on the occasion of his 70th birthday}
\end{center}
\end{titlepage}
\section{Introduction and results}
The dynamics of the D-branes of type II superstring theories is well-approximated by the effective world-volume field theories which consist of the sum of   Dirac-Born-Infeld (DBI) and Chern-Simons (CS) actions. 
The DBI action  describes the dynamics of the brane in the presence of the NSNS background fields. For constant background fields  it can be found by requiring the consistency with  nonlinear T-duality \cite{Leigh:1989jq,Bachas:1995kx}
\beqa
S_{DBI}&=&-T_p\int d^{p+1}x\,e^{-\phi}\sqrt{-\det\left(G_{ab}+B_{ab}+2\pi\alpha'F_{ab}\right)}
\eeqa
where $G_{ab}$ and $B_{ab}$ are  the pulled back of the bulk fields $G_{\mu\nu}$ and $B_{\mu\nu}$ onto the world-volume of D-brane\footnote{Our index conversion is that the Greek letters  $(\mu,\nu,\cdots)$ are  the indices of the space-time coordinates, the Latin letters $(a,d,c,\cdots)$ are the world-volume indices and the letters $(i,j,k,\cdots)$ are the normal bundle indices.}. The curvature corrections to this action has been found in \cite{Bachas:1999um} by requiring consistency of the effective action with the $O(\alpha'^2)$ terms of the corresponding disk-level scattering amplitude \cite{Garousi:1996ad,Hashimoto:1996kf}. The on-shell ambiguity of these couplings  has been removed in \cite{Garousi:2009dj} by requiring the consistency of the couplings with linear T-duality. Moreover, this consistency fixes the couplings of non-constant dilaton and B-field at the order  $O(\alpha'^2)$  in the action which are reproduced by the corresponding disk level scattering amplitude.  In particular, it has been found in  \cite{Garousi:2009dj} that the consistency with T-duality/S-matrix requires the non-constant dilaton appears in the string frame action only as the overall factor of $e^{-\phi}$.  
%For totally-geodesic embeddings of world-volume in the ambient spacetime, the corrections in string frame  for zero $B_{\mu\nu},F_{ab}$ and for constant dilaton  are \cite{Bachas:1999um}
%\beqa
%S&\!\!\!\!=\!\!\!\!&\frac{T_p}{4}\frac{(4\pi^2\alpha')^2}{32\pi^2}\int d^{p+1}x\,e^{-\phi}\sqrt{-G}\left(R_{abcd}R^{abcd}-2\hR_{ab}\hR^{ab}-R_{abij}R^{abij}+2\hR_{ij}\hR^{ij}\right)\labell{DBI}
%\eeqa
%where $\hR_{ab}=G^{cd}R_{cadb}$ and $\hR_{ij}=G^{cd}R_{cidj}$. Here also a tensor with the world-volume or transverse space indices is the pulled back of the corresponding  bulk tensor onto world-volume or transverse space. For the case of D$_3$-brane with trivial normal bundle the above curvature couplings have been modified in \cite{Bachas:1999um} to include the complete sum of D-instanton corrections by  requirement the $SL(2,Z)$ invariance of the couplings.

The CS part on the other hand describes the coupling of D-branes to the RR fields. For constant background fields it is given by \cite{Polchinski:1995mt,Douglas:1995bn}
\beqa
S_{CS}&=&T_{p}\int_{M^{p+1}}e^{B}C\labell{CS2}
\eeqa
where $M^{p+1}$ represents the world volume of the D$_p$-brane, $C$ is meant to represent a sum over all appropriate RR potential forms  and the multiplication rule is the wedge product. The abelian gauge field can be added to the action as $B\rightarrow B+2\pi\alpha'F$. The curvature corrections to this action has been found 
 by requiring that the chiral anomaly on the world volume of intersecting D-branes (I-brane) cancels with the anomalous variation of the CS action \cite{Green:1996dd,Cheung:1997az,Minasian:1997mm}. This correction is
\beqa
S_{CS}&=&T_{p}\int_{M^{p+1}}e^BC\left(\frac{{\cal A}(4\pi^2\alpha'R_T)}{{\cal A}(4\pi^2\alpha'R_N)}\right)^{1/2}\labell{CS}
\eeqa
where  ${\cal A}(R_{T,N})$ is the Dirac roof genus of the tangent and normal bundle curvatures respectively,
\beqa
\sqrt{\frac{{\cal A}(4\pi^2\alpha'R_T)}{{\cal A}(4\pi^2\alpha'R_N)}}&=&1+\frac{(4\pi^2\alpha')^2}{384\pi^2}(\tr R_T^2-\tr R_N^2)+\cdots \labell{roof}
\eeqa
For totally-geodesic embeddings of world-volume in the ambient spacetime,   ${\rm R}_{T,N}$ are the pulled back curvature 2-forms of the tangent and normal bundles respectively (see the appendix in ref.
\cite{Bachas:1999um} for more details).

It has been pointed out in \cite{Myers:1999ps} that the anomalous CS couplings \reef{CS} must be incomplete for non-constant B-field as they are not compatible with T-duality. T-duality exchanges the components of the metric and  the B-field whereas  the couplings \reef{CS} includes only the curvature terms. Compatibility of this action  with T-duality should give a bunch of new couplings \cite{Ga, kg}.
%It has been speculated in \cite{...} that the consistent couplings may involve the generalized curvatures of a torsionful connection constructed by combining the metric spin-connection and the NS-NS three-form field strength. 

In this paper we would like to show that for non-constant RR and NSNS fields there  are other contribution to the action \reef{CS} at order $O(\alpha'^2)$ which may not  arise from requiring the consistency of the action \reef{CS} with T-duality. These terms which involve linear NSNS field can be found by studying the S-matrix element of one RR and one NSNS vertex operators  \cite{Garousi:1996ad} and by requiring them to be consistent  with linear T-duality. We will find  the following string frame couplings at order $O(\alpha'^2)$:
\beqa
S_{}&\sim&T_p\int d^{p+1}x\,\eps^{a_0\cdots a_p}\left(\frac{1}{2!(p-1)!}[F^{(p)}_{ia_2\cdots a_p,a}H_{a_0a_1}{}^{a,i}-\frac{1}{p}F^{(p)}_{a_1a_2\cdots a_p,i}(H_{a_0a}{}^{i,a}-H_{a_0j}{}^{i,j})]\right.\nonumber\\
&&\left.\qquad\qquad\qquad\qquad+\frac{2}{p!}[\frac{1}{2!}F^{(p+2)}_{ia_1\cdots a_pj,a}\cR^a{}_{a_0}{}^{ij}-\frac{1}{p+1}F^{(p+2)}_{a_0\cdots a_pj,i}\hat{\cR}^{ij}]\right.\nonumber\\
&&\left.\qquad\qquad\qquad\qquad-\frac{1}{3!(p+1)!}F^{(p+4)}_{ia_0\cdots a_pjk,a}H^{ijk,a}\right)\labell{LTdual}
\eeqa
where as usual commas denote partial differentiation. 

It has been shown in \cite{Garousi:2009dj} that the compatibility of the curvature corrections to the DBI action with linear T-duality transformations 
%and consistency with the S-matrix elements 
requires the non-constant dilaton appears in the string frame action only through  the overall factor of $e^{-\phi}$. This factor has been absorbed in the RR field so one expects that the dilaton appears in the above action only through the string frame metric. We will show that the coupling of one $F^{(p+2)}$ and one dilaton in the Einstein frame which can be calculated by the S-matrix element, is reproduced exactly by transforming the   couplings  in the second line above to the Einstein frame. 

The couplings in \reef{LTdual} have been found by the S-matrix element of one RR and one NSNS vertex operators and by T-duality. The S-matrix method  produces the on-shell couplings and consistency of the couplings with linear T-duality then  fixes the on-shell ambiguity of the couplings. Correction to this action can also be found  by requiring it to be consistent with nonlinear T-duality transformations.   We will consider one particular nonlinear term in the T-duality transformation of the RR field and then examine the consistency of \reef{LTdual} with it to find new couplings. The new couplings  are given by the above action in which $F^{(n)}$ is replaced by ${\cal F}^{(n)}$ where
\beqa
{\cal F}^{(n)}&=&F^{(n)}+B\wedge F^{(n-2)}+H\wedge C^{(n-3)}\labell{newF}\\
&&+\frac{1}{2!}B\wedge B\wedge F^{(n-4)}+\frac{1}{2!}B\wedge H\wedge  C^{(n-5)}+\frac{1}{2!}H\wedge B\wedge C^{(n-5)}+\cdots\nonumber\\
&=&d{\cal C}^{(n)}\nonumber
\eeqa
where  ${\cal C}=e^{B}C$, is the RR potential in the CS action \reef{CS2}.

An outline of the paper is as follows: We begin the section 2 by writing the S-matrix element of one RR and one NSNS vertex operators. From the contact terms of this amplitude at order $(\alpha')^2$, we will find the on-shell couplings of one massless RR and two NSNS fields. In section 3, we review the T-duality transformations and the  strategy for checking the consistency of a D-brane action  with T-duality. In section 3.1, we check the consistency of the couplings found in section 2 with linear T-duality which fixes the on-shell ambiguity of the couplings. After fixing the on-shell ambiguity of the gravity couplings, we show that the dilaton appears in the action only through the string  frame metric. In section 3.2, we check the consistency of the couplings \reef{LTdual} with nonlinear T-duality and show that the field strength in the action \reef{LTdual} should be given by \reef{newF}. 

\section{Scattering amplitudes}

A  method for finding the couplings in effective field theory is the S-matrix method. The standard CS coupling \reef{CS2} has been confirmed by the S-matrix method in \eg \cite{Garousi:1998bj,Dymarsky:2002kk}.  The couplings of NSNS and  RR fluxes to various types of D-branes  have been found in \cite{Billo':2008sp} by evaluating disk amplitudes among two open string and one closed string vertex operators. To find the couplings of one RR and one NSNS states to D$_p$-brane, one needs 
the scattering amplitude of their corresponding vertex operators which  is given by \cite{Garousi:1996ad}
 \beqa
 A(\veps_1,p_1;\veps_2,p_2)&=&-\frac{1}{8}T_p\alpha'^2K(1,2)\frac{\Gamma(-\alpha' t/4)\Gamma(\alpha' q^2)}{\Gamma(1-\alpha't/4+\alpha' q^2)}\nonumber\\
 &=&\frac{1}{2}T_pK(1,2)\left(\frac{1}{q^2t}+\frac{\pi^2\alpha'^2}{24}+O(\alpha'^4)\right)\labell{Amp}
 \eeqa
where $q^2=p_1^ap_1^b\eta_{ab}$ is the momentum flowing along the world-volume of D-brane, and $t=-(p_1+p_2)^2$ is the momentum transfer in the transverse direction. The kinematic factor is
\beqa
K(1,2)&=&i\frac{q^2}{\sqrt{2}}\Tr(P_-\Gamma_{1(n)}M_p\gamma^{\nu}\gamma\inn(p_1+p_2)\gamma^{\mu})(\veps_2\inn D)_{\mu\nu}\labell{kin}\\
&&-i\frac{t}{2\sqrt{2}}[\Tr(P_-\Gamma_{1(n)}M_p\gamma\inn D\inn\veps_2^T\inn D\inn p_2)-\Tr(P_-\Gamma_{1(n)}M_p\gamma\inn\veps_2\inn D\inn p_2)\nonumber\\
&&\qquad\qquad\qquad\qquad\qquad\qquad\qquad\qquad-\Tr(P_-\Gamma_{1(n)}M_p\gamma\inn D\inn p_2)\Tr(\veps_2\inn D)]\nonumber
\eeqa
where  the matrix $D^{\mu}_{\nu}$ is diagonal with +1 in the world volume directions and -1 in the transverse directions, and
\beqa
\Gamma_{1(n)}&=&\frac{1}{n!}F_{1\nu_1\cdots\nu_n}\gamma^{\nu_1}\cdots\gamma^{\nu_n}\nonumber\\
M_p&=&\frac{\pm 1}{(p+1)!}\eps_{a_0\cdots a_p}\gamma^{a_0}\cdots\gamma^{a_p}
\eeqa
where $F_1$ is the linearized RR field strength $n$-form and $\eps$ is the volume $p+1$-form of the $D_p$-brane. In equation \reef{kin}, $P_-=\frac{1}{2}(1-\gamma_{11})$ is the chiral projection operator and $\veps_2$ is  the NS-NS polarization. The $\gamma_{11}$ in the chiral projection gives the magnetic couplings and $1$ gives the electric couplings. The first term in \reef{Amp} produces the massless poles resulting from  the $(\alpha')^0$ order of the DBI and CS couplings on the D-brane, and  the supergravity couplings in the bulk.  The second term in \reef{Amp} should produce $(\alpha')^2$ couplings of one RR and one NSNS on the D-brane in which we are interested.

Using the identity $M_p\gamma^{\mu}=D^{\mu}{}_{\nu}\gamma^{\nu}M_p$ \cite{Garousi:1996ad} and the algebra $\{\gamma^{\mu},\gamma^{\nu}\}=-2\eta^{\mu\nu}$, one can write the above kinematic factor for the electric couplings as
\beqa
K(1,2)&\!\!\!=\!\!\!&i\frac{q^2}{2\sqrt{2}}\Tr(\Gamma_{1(n)}\gamma^{\nu}M_p\gamma\inn(p_1+p_2)\gamma^{\mu})(\veps_2)_{\mu\nu}\labell{kin2}\\
&&-i\frac{t}{4\sqrt{2}}[%\Tr(\Gamma_{1(n)}\gamma\inn(\veps_2^T-\veps_2)\inn D\inn p_2M_p)
-\Tr(\Gamma_{1(n)}\gamma\inn D\inn\veps_2\inn D\inn p_2M_p)\nonumber\\
&&-\frac{1}{2}\Tr(\Gamma_{1(n)}\gamma^{\mu}M_p\gamma\inn D\inn p_2\gamma^{\nu})(D\inn\veps_2\inn D+\veps^T_2)_{\mu\nu}]\nonumber
\eeqa
One can easily check that the kinematic factor is zero for $n\le p-2$ and for $n> p+4$. This factor is non-zero for $n=p,\, n=p+2$ and for $n=p+4$. Let us consider each case separately.

\subsection{$n=p$ case}

For $n=p$ case,  one needs to perform the following traces:
\beqa
\Tr(\ga^{\mu_1}\cdots\ga^{\mu_p}\ga^{\mu}\ga^{a_0}\cdots\ga^{a_p}\ga^{\alpha}\ga^{\nu})&{\rm and}& \Tr(\ga^{\mu_1}\cdots\ga^{\mu_p}\ga^{\alpha}\ga^{a_0}\cdots\ga^{a_p})
\eeqa
They  make  various contraction of the indices. The  first one simplifies to
\beqa
&&(-1)^p[\eta^{\mu\nu}\eta^{\alpha a_0}\eta^{\mu_1a_1}+\eta^{\alpha\nu}\eta^{a_0\mu}\eta^{a_1\mu_1}-\eta^{\alpha\mu}\eta^{a_0\nu}\eta^{a_1\mu_1}-p\eta^{\alpha\mu_1}\eta^{a_0\mu}\eta^{a_1\nu}\nonumber\\
&&-p\eta^{\alpha a_0}\eta^{\mu\mu_1}\eta^{a_1\nu}-p\eta^{\alpha a_0}\eta^{a_1\mu}\eta^{\nu\mu_1}]
 p(p+1)\Tr(\ga^{\mu_2}\cdots\ga^{\mu_p}\ga^{a_2}\cdots\ga^{a_p})
\eeqa
The second trace simplifies to 
\beqa
\Tr(\ga^{\mu_1}\cdots\ga^{\mu_p}\ga^{\alpha}\ga^{a_0}\cdots\ga^{a_p})&=&(-1)^p\eta^{\alpha a_0}\eta^{\mu_1 a_1}p(p+1)\Tr(\ga^{\mu_2}\cdots\ga^{\mu_p}\ga^{a_2}\cdots\ga^{a_p})
\eeqa
and the trace $\Tr(\ga^{\mu_2}\cdots\ga^{\mu_p}\ga^{a_2}\cdots\ga^{a_p})$ causes the RR field strength to contract with the volume form $\eps$, \ie $F_{1\mu_1a_2\cdots a_p}\eps^{a_0\cdots a_p}$.

Using the above traces, one finds that the kinematic factor \reef{kin2}  for the graviton is
\beqa
K(1,2)&\sim&-i\frac{t}{2\sqrt{2}}[F_{1a_1\cdots a_p}(\veps_2)_{a_0}{}^ap_{2a}+pF_{1aa_2\cdots a_p}(\veps_2)_{a_1}{}^ap_{2a_0}]\eps^{a_0\cdots a_p}\labell{K12}
\eeqa
Using the fact that the indices $a_0,\cdots, a_p$ contracted with the totally antisymmetric  $\eps^{a_0\cdots a_p}$ tenser and the conservation of the momentum $p_{1a}+p_{2a}=0$, one can write 
\beqa
pF_{1aa_2\cdots a_p}p_{2a_0}&=&p\,p_{1a}C_{1a_2\cdots a_p}p_{2a_0} \nonumber\\
&=&F_{1a_0a_2\cdots a_p}p_{2a}\labell{iden1}
\eeqa
which makes the kinematic factor \reef{K12} to be zero. For B-field, one finds the following non-zero result for the kinematic factor \reef{kin2}: 
\beqa
K(1,2)&\sim &-\left(i\frac{q^2}{2\sqrt{2}}[2F_{1a_1a_2\cdots a_p}(\veps_2)_{a_0i}(p_1+p_2)^i-pF_{1ia_2\cdots a_p}(\veps_2)_{a_0a_1}(p_1+p_2)^i]\right.\nonumber\\
&&\left.-i\frac{t}{\sqrt{2}}F_{1a_1a_2\cdots a_p}(\veps_2)_{a_0i}p_2^i\right)\eps^{a_0a_1\cdots a_p}
\eeqa
 As a check of the calculation, if one replaces the B-field polarization with $(\veps_2)_{\mu\nu}\rightarrow \z_{\mu}(p_2)_\nu-\z_{\nu}(p_2)_\mu$  the kinematic factor vanishes, as expected from the Ward identity.
 
 To find the field theory couplings corresponding to the above momentum space contact terms, we use the following identities:
\beqa
F_{1a_1\cdots a_p}(\veps_2)_{a_0a}&=&\frac{p}{2}F_{1aa_2\cdots a_p}(\veps_2)_{a_0a_1}\nonumber\\
F_{1a_1\cdots a_p}(p_1)_i&=&pF_{1ia_2\cdots a_p}(p_1)_{a_1}
\eeqa
where we have used the fact that the indices $a_0,\cdots, a_p$ contracted with the totally antisymmetric  $\eps^{a_0\cdots a_p}$ tenser. Using these identities one can write the kinematic factor as
\beqa
K(1,2)&\sim &-\frac{ip}{\sqrt{2}}\left((\veps_2)_{a_0a_1}[-\frac{1}{2}p_1\inn V\inn p_1F_{1ia_2\cdots a_p}(p_2)^i-p_1\inn N\inn p_2F_{1aa_2\cdots a_p}(p_2)^a]\right.\nonumber\\
&&\left.+p_1\inn V\inn p_1F_{1ia_2\cdots a_p}(p_1)_{a_1}(\veps_2)_{a_0}{}^i\right)\eps^{a_0a_1\cdots a_p}
\eeqa 
which satisfies the Ward identity. The couplings corresponding to the above terms are:
\beqa
\frac{T_p}{2!(p-1)!}\int d^{p+1}x\,\eps^{a_0a_1\cdots a_p}\left(F^{(p)}_{ia_2\cdots a_p,a}H_{a_0a_1}{}^{a,i}- F^{(p)}_{aa_2\cdots a_p,i}H_{a_0a_1}{}^{i,a}\right)\labell{first}
\eeqa
%Since the index $i$ is a derivative index, the T-duality in a transverse coordinate gives no couplings involving $F^{(p-2)}$ which is consistent with the S-matrix element.
where 
\beqa
H_{\mu\nu\alpha}&=&B_{\mu\nu,\alpha}+B_{\alpha\mu,\nu}+B_{\nu\alpha,\mu}
\eeqa
The other terms in \reef{Amp} correspond to the higher derivative of the couplings \reef{first} in which we are not interested in this paper. 

The last coupling in \reef{first} has on-shell ambiguity. To see this we note that the index $a$ in this term can be either $a_0$ or $a_1$. If $a=a_0$, it can be written as $-2F_{a_0a_2\cdots a_p,i}H_{aa_1}{}^{i,a}/p$, and if $a=a_1$, it  can be written as $-2F_{a_1a_2\cdots a_p,i}H_{a_0a}{}^{i,a}/p$. Interchanging $a_1\leftrightarrow a_0$ in the latter  expression and using the fact that it has the overall factor of the volume form, one can write it as the former expression. Hence, the last term in \reef{first} can be written as $-2F_{a_1a_2\cdots a_p,i}H_{a_0a}{}^{i,a}/p$. Moreover, using the on-shell condition $H_{\nu\rho\alpha,\mu}{}^{\mu}=0$, one can write it as $2F_{a_1a_2\cdots a_p,i}H_{a_0j}{}^{i,j}/p$ or as
\beqa
-\frac{1}{p}F_{a_1\cdots a_p,i}(H_{a_0a}{}^{i,a}-H_{a_0j}{}^{i,j})
\eeqa
We will fix the above on-shell ambiguity in section 3 by requiring the consistency of the coupling with the T-duality transformations.

\subsection{$n=p+2$ case}

For $n=p+2$ case, the traces in \reef{kin2} simplify to
\beqa
\Tr(\ga^{\mu_1}\cdots\ga^{\mu_{p+2}}\ga^{\mu}\ga^{a_0}\cdots\ga^{a_p}\ga^{\alpha}\ga^{\nu})&\!\!\!=\!\!\!&(p+1)(p+2)\Tr(\ga^{\mu_3}\cdots\ga^{\mu_{p+2}}\ga^{a_1}\cdots\ga^{a_p})\times\nonumber\\
&&[-\eta^{\mu\nu}\eta^{\alpha\mu_1}\eta^{a_0\mu_2}+\eta^{\alpha\nu}\eta^{a_0\mu_2}\eta^{\mu\mu_1}+\eta^{\alpha\mu}\eta^{\mu_1\nu}\eta^{a_0\mu_2}+\nonumber\\
&&(p+1)(\eta^{\alpha\mu_1}\eta^{\mu_2\mu}\eta^{a_0\nu}+\eta^{\alpha \mu_1}\eta^{\mu a_0}\eta^{\mu_2\nu}+\eta^{\alpha a_0}\eta^{\mu_1\mu}\eta^{\nu\mu_2})]\nonumber\\ 
\Tr(\ga^{\mu_1}\cdots\ga^{\mu_{p+2}}\ga^{\alpha}\ga^{a_0}\cdots\ga^{a_p})&=&-\eta^{\alpha \mu_1}\eta^{\mu_2 a_0}(p+1)(p+2)\Tr(\ga^{\mu_3}\cdots\ga^{\mu_{p+2}}\ga^{a_1}\cdots\ga^{a_p})\nonumber
\eeqa
The trace $\Tr(\ga^{\mu_3}\cdots\ga^{\mu_{p+2}}\ga^{a_1}\cdots\ga^{a_p})$ causes the RR field strength to contract with the volume form as $F_{i\mu_1\mu_2a_1\cdots a_p}\eps^{a_0\cdots a_p}$. Inserting these traces in \reef{kin2}, one finds
the kinematic factor  for B-field becomes
\beqa
K(1,2)&\sim & -i\frac{t}{2\sqrt{2}}[-F_{1ia_0\cdots a_p}(\veps_2)^i{}_a p_{2a}+(p+1)F_{1iaa_1\cdots a_p}(\veps_2)^i{}_ap_{2a_0}]\eps^{a_0\cdots a_p}\labell{K122}
\eeqa
Using the fact that the indices $a_0,\cdots, a_p$ contract with the totally antisymmetric  tensor $\eps^{a_0\cdots a_p}$ and the conservation of the momentum $p_{1a}+p_{2a}=0$, one can write 
\beqa
(p+1)F_{1iaa_1\cdots a_p}p_{2a_0}\ell^a
%&=&(p+1)(p_{1i}C_{1aa_1\cdots a_p}-p_{1a}C_{1ia_2\cdots a_p})p_{2a_0}\ell^a \nonumber\\
%&=&[p_{1i}C_{1a_0a_1\cdots a_p}-(p+1)p_{1a_0}C_{1ia_2\cdots a_p}]p_{2a}\ell^a \nonumber\\
&=&F_{1ia_0a_1\cdots a_p}p_{2a}\ell^a\labell{iden2}
\eeqa
for any vector $\ell^a$. This  makes the kinematic factor \reef{K122} to be zero. For the symmetric polarization, graviton or dilaton, one finds the kinematic factor \reef{kin2} to be 
\beqa
K(1,2)&\sim &\left(i\frac{q^2}{\sqrt{2}}[F_{1ja_0\cdots a_p}(\veps_2)^j{}_i(p_1+p_2)^i+(p+1)F_{1i\mu a_1\cdots a_p}(\veps_2)^{\mu}{}_{a_0}(p_1+p_2)^i]\right.\nonumber\\
&&+i\frac{t}{2\sqrt{2}}[2F_{1ia_0\cdots a_p}(\veps_2)^{ia}p_{2a}-(p+1)F_{1aa_1\cdots a_p}(\veps_2)^a{}_{a_0}p_2^i]\nonumber\\
&&\left.-\frac{i}{2\sqrt{2}}F_{1ia_0\cdots a_p}p_2^i(q^2-\frac{t}{2})\Tr(\veps_2)\right)\eps^{a_0a_1\cdots a_p}\labell{kin3}
\eeqa
The last term is zero for graviton, but is has contribution to the dilaton amplitude. Using the identity \reef{iden2}, one can write the above equation for the graviton as
\beqa
K(1,2)&\sim &\frac{i(p+1)}{\sqrt{2}}\left(F_{1jaa_1\cdots a_p}p_{2a_0}p_2^a(\veps_2)^j{}_ip_1^i-p_1\inn V\inn p_2F_{1ija_1\cdots a_p}(\veps_2)^{j}{}_{a_0}p_2^i\right.\nonumber\\
&&\left.+p_1\inn N\inn p_2[F_{1iaa_1\cdots a_p}(\veps_2)^{a}{}_{a_0}p_{2}^i-2F_{1iaa_1\cdots a_p}(\veps_2)^{ai}p_{2a_0}]\right)\eps^{a_0a_1\cdots a_p}
\eeqa
It satisfies the Ward identity. Using the identity 
\beqa
F_{1jaa_1\cdots a_p}p_{1i}p_{2a_0}&=&F_{1iaa_1\cdots a_p}p_{1j}p_{2a_0}-F_{1ija_1\cdots a_p}p_{1a}p_{2a_0}\,,
\eeqa
one finds that the field theory corresponding to the above amplitude is
\beqa
\frac{T_p}{p!}\int d^{p+1}x\,\eps^{a_0a_1\cdots a_p}\left(\frac{1}{2}F^{(p+2)}_{ija_1\cdots a_p,a}\cR^a{}_{a_0}{}^{ij}+F^{(p+2)}_{jaa_1\cdots a_p,i}\cR^{i}{}_{a_0}{}^{aj}\right)\labell{second}
\eeqa
The Riemann tensor at the linear order in the graviton is 
\beqa
\cR_{\mu\nu\rho\la}&=&\frac{1}{2}(h_{\mu\la,\nu\rho}+h_{\nu\rho,\mu\la}-h_{\mu\rho,\nu\la}-h_{\nu\la,\mu\rho})
\eeqa
where we have considered perturbation around the flat space where the metric takes the form $G_{\mu\nu}=\eta_{\mu\nu}+h_{\mu\nu}$. The last term in the above amplitude has again on-shell ambiguity. We will show in section 3 that this term in the present form is not consistent with T-duality. However, it can be written in a T-dual invariant form using the on-shell conditions.

The  dilaton amplitude can be found from the amplitude \reef{kin3}  by using the following  polarization:
\beqa
\veps_{\mu\nu}&=&\frac{1}{\sqrt{8}}(\eta_{\mu\nu}-\ell_{\mu}p_{\nu}-\ell_{\nu}p_{\mu})\,;\qquad \ell\inn p=1\labell{pol}
\eeqa
where the auxiliary vector  $\ell_{\mu}$ insures that the polarization satisfies the on-shell condition $p\inn\veps_{\nu}=0$. One finds the dilaton amplitude to be
\beqa
K(1,2)&\sim &\frac{i(p-3)}{\sqrt{2}}\left(p_1\inn N\inn p_2 F_{1ia_0\cdots a_p}p_2^i\right)\eps^{a_0a_1\cdots a_p}
\eeqa
The field theory corresponding to the above amplitude is
\beqa
\frac{(p-3)T_p}{(p+1)!}\int d^{p+1}x\,\eps^{a_0a_1\cdots a_p}\left(F^{(p+2)}_{ia_0\cdots a_p,j}\phi\,^{,ij}\right)\labell{dil}
\eeqa
This coupling is zero for D$_{3}$-brane which is consistent with the fact that the world volume theory of D$_{3}$-brane is  a conformal field theory.

\subsection{$n=p+4$ case}

For $n=p+4$ case, one needs only the following trace:
\beqa
&&\Tr(\ga^{\mu_1}\cdots\ga^{\mu_{p+4}}\ga^{\mu}\ga^{a_0}\cdots\ga^{a_p}\ga^{\alpha}\ga^{\nu})\nonumber\\
&&=(-1)^p\eta^{\alpha \mu_1}\eta^{\mu\mu_2 }\eta^{\nu\mu_3}(p+2)(p+3)(p+4)\Tr(\ga^{\mu_4}\cdots\ga^{\mu_{p+4}}\ga^{a_0}\cdots\ga^{a_p})
\eeqa
One can easily check  that the kinematic factor \reef{kin2}  is zero for graviton, and for B-field it is
\beqa
K(1,2)&\sim & -i\frac{q^2}{2\sqrt{2}}F_{1ijka_0\cdots a_p}(\veps_2)^{jk}p_2^i\eps^{a_0\cdots a_p}\labell{K123}
\eeqa
which  satisfies the Ward identity. The coupling  corresponding to the above amplitude is 
\beqa
\frac{T_p}{3!(p+1)!}\int d^{p+1}x\,\eps^{a_0\cdots a_p}F^{(p+4)}_{ijka_0\cdots a_p,a}H^{ijk,a}\labell{third}
\eeqa
Note that $H^{ijk,a}=H^{ija,k}$ when the indices $i,j,k$ are totally antisymmetric, as in above equation. In the next section, we will examine  that  the  consistency of the above   couplings   with T-duality. 
 
\section{T-duality}

In this section  we would like to study the transformation of the couplings that we have found in the previous section under the T-duality. We denote $y$ the Killing direction along which the  T-duality is going to be implemented. 
The full set of T-duality transformations has been found in \cite{TB,Meessen:1998qm}
\beqa
e^{2\tphi}&=&\frac{e^{2\phi}}{G_{yy}}\nonumber\\
\tG_{yy}&=&\frac{1}{G_{yy}}\nonumber\\
\tG_{\mu y}&=&\frac{B_{\mu y}}{G_{yy}}\nonumber\\
\tG_{\mu\nu}&=&G_{\mu\nu}-\frac{G_{\mu y}G_{\nu y}-B_{\mu y}B_{\nu y}}{G_{yy}}\nonumber\\
\tB_{\mu y}&=&\frac{G_{\mu y}}{G_{yy}}\nonumber\\
\tB_{\mu\nu}&=&B_{\mu\nu}-\frac{B_{\mu y}G_{\nu y}-G_{\mu y}B_{\nu y}}{G_{yy}}\nonumber\\
\tC^{(n)}_{\mu\cdots \nu\alpha y}&=&C^{(n-1)}_{\mu\cdots \nu\alpha }-(n-1)\frac{C^{(n-1)}_{[\mu\cdots\nu|y}G^{}_{|\alpha]y}}{G_{yy}}\nonumber\\
\tC^{(n)}_{\mu\cdots\nu\alpha\beta}&=&C^{(n+1)}_{\mu\cdots\nu\alpha\beta y}+nC^{(n-1)}_{[\mu\cdots\nu\alpha}B^{}_{\beta]y}+n(n-1)\frac{C^{(n-1)}_{[\mu\cdots\nu|y}B^{}_{|\alpha|y}G^{}_{|\beta]y}}{G_{yy}}\labell{Cy}
\eeqa
where $\mu,\nu,\alpha,\beta\ne y$. In above transformation the metric is the string frame metric. If $y$ is identified on a circle of radius $R$, \ie $y\sim y+2\pi R$, then after T-duality the radius becomes $\tilde{R}=\alpha'/R$. The string coupling is also shifted as $\tilde{g}=g\sqrt{\alpha'}/R$.
 % The dilaton transformation indicates that if the dilaton is constant in the original theory, it is not so in the T-dual theory, unless one assumes that $h_{yy}$ in the original theory to be a  constant. 
 
 The strategy for finding T-duality invariant couplings  is given in \cite{Garousi:2009dj}. Let us review it here.  
Suppose we are implementing T-duality along a world volume direction of a D$_p$-brane denoted $y$. One should first separate the  world-volume indices   into  $y$ index and the world-volume indices which do not include $y$, and then apply the above T-duality transformations.  The  latter indices  are  complete world-volume indices   of the  T-dual D$_{p-1}$-brane. However, the $y$ index in the T-dual theory which is a normal bundle index is not a complete index. 
%One must then include some other terms in the original theory to have totally completed indices in the T-dual theory. 
On the other hand, the normal bundle indices of  the original theory  are not complete in the T-dual D$_{p-1}$-brane. They are not include $y$.  In a T-duality invariant theory, the $y$ indices  must be combined  with the incomplete normal bundle indices  to give the complete normal bundle indices in the T-dual D$_{p-1}$-brane.  If a theory is not invariant under the T-duality,  one should then add  new terms  to it  to have   the complete indices in the T-dual theory. In this way one makes the theory to be T-duality invariant by adding new couplings.

One may also  implement T-duality along a transverse direction of a D$_p$-brane denoted $y$. In this case one must first separate the transverse indices to $y$ and the transverse indices  which do not include $y$, and then apply the above T-duality transformations. The latter indices  are  the complete transverse  indices   of the  T-dual D$_{p+1}$-brane. However, the  complete world-volume indices of the original D$_p$-brane    are not the complete indices of the T-dual D$_{p+1}$-brane. They must  include the $y$ index to be complete.  In a T-duality invariant theory, the $y$ index which is a world-volume index in the T-dual theory must be combined with the incomplete world-volume indices  of the T-dual D$_{p+1}$-brane to give the complete world-volume indices.

\subsection{Linear T-duality }

In this subsection we would like to study the consistency of  the couplings  with linear T-duality transformations. Assuming the NSNS and RR fields are small perturbations around the flat space, the T-duality  transformations take the following linear form:
\beqa
&&\tilde{\phi}=\phi-\frac{1}{2}{h}_{yy},\,
\tilde{h}_{yy}=-h_{yy},\, \tilde{h}_{\mu y}=B_{\mu y},\, \tilde{B}_{\mu y}=h_{\mu y},\,\tilde{h}_{\mu\nu}=h_{\mu\nu},\,\tilde{B}_{\mu\nu}=B_{\mu\nu}\nonumber\\
&&\tC^{(n)}_{\mu\cdots \nu\alpha y}=C^{(n-1)}_{\mu\cdots \nu\alpha },\,\tC^{(n)}_{\mu\cdots\nu\alpha\beta}=C^{(n+1)}_{\mu\cdots\nu\alpha\beta y}\labell{linear}
\eeqa
Consistency of the curvature squared corrections to the DBI action under the above linear T-duality transformations has been examined in \cite{Garousi:2009dj}. The consistency requires adding some $H$-squared terms to the DBI action which are also consistent with the corresponding S-matrix element. We are going to do similar calculation for the couplings that we have found in the previous section.  

We begin by studying the T-duality of the couplings in \reef{first}. Consider implementing T-duality along a world volume direction of the D$_p$-brane\footnote{The couplings \reef{first} are consistent with the linear T-duality transformations \reef{linear} when  implementing the T-duality along a transverse direction.}.   From the contraction with the world volume form, one of the indices $a_2,\cdots, a_{p}$ of the RR field strength or the indices  $a_0,a_1$ of the NSNS field strength  in \reef{first} must include $y$, and so there are two cases to consider: First when $y$ appears as an index on the RR field strength and second when $y$ is an index in the NSNS field strength. In the former case, we  note that there are $p-1$ indices in the RR field strength which are contracted with the volume form. Each of these indices can be $y$. However, because of the totally antisymmetric property of the volume form and the RR field strength, they all are identical. So one can write \reef{first} as
\beqa
\frac{T_p}{2!(p-2)!}\int d^{p+1}x\,\eps^{a_0a_1\cdots a_{p-1}y}\left(F^{(p)}_{ia_2\cdots a_{p-1}y,a}H_{a_0a_1}{}^{a,i}- F^{(p)}_{aa_2\cdots a_{p-1}y,i}H_{a_0a_1}{}^{i,a}\right)\labell{first1}
\eeqa
 Note that the indices $i,a$ appear as the derivative indices so nigher of them can be $y$. Moreover, because of the world volume form, none of the indices $a_0,\cdots a_{p-1}$ can be $y$. The  transform of the above couplings under the linear T-duality \reef{linear} gives  the following couplings for D$_{p-1}$-brane:
\beqa
2\pi\sqrt{\alpha'}\frac{T_p}{2!(p-2)!}\int d^{p}x\,\eps^{a_0a_1\cdots a_{p-1}}\left(F^{(p-1)}_{ia_2\cdots a_{p-1},a}H_{a_0a_1}{}^{a,i}- F^{(p-1)}_{aa_2\cdots a_{p-1},i}H_{a_0a_1}{}^{i,a}\right)\nonumber
\eeqa
where we have also used the fact that $T_p\sim 1/g_s$. Using the relation $2\pi\sqrt{\alpha'}T_p=T_{p-1}$, one observes that the above couplings  are exactly the couplings \reef{first} for D$_{p-1}$-brane. 

%In order to proceed further, one observes that in the action \reef{first} there are two  indices carried by NSNS field strength  namely $a_0,a_1$ which  are contracted with the volume form.  When performing T-duality along a particular world volume direction either one of these or one of the indices on the RR field strength must equal the T-dual coordinate $y$. We already showed that  the case in which the index $y$ was carried by the RR  field strength is consistent with T-duality and

We  will now check the case that the T-dual coordinate $y$ is carried by the NSNS field strength. 
%The strategy is to choose one of the two indices to perform the T-duality on and infer what extra terms must be included for the consistency. The resulting terms will have one remaining indices and we will then choose it  to perform T-duality on and infer again which terms must be present to maintain consistency. 
There are two    possibilities for the NSNS field strength  in \reef{first} to carry the T-dual coordinate $y$, \ie either $a_0$ or $a_1$ carries the index $y$.  Since the two possibilities are identical, one can write \reef{first} as
\beqa
\frac{T_p}{(p-1)!}\int d^{p+1}x\,\eps^{ya_1\cdots a_{p}}\left(F^{(p)}_{ia_2\cdots a_{p},a}H_{ya_1}{}^{a,i}- F^{(p)}_{aa_2\cdots a_{p},i}H_{ya_1}{}^{i,a}\right)\labell{first2}
\eeqa
Note again that the indices $i,a$ and $a_1,\cdots , a_p$ can not be $y$. The above couplings   transform under linear T-duality to the following couplings of D$_{p-1}$-brane:
\beqa
\frac{T_{p-1}}{(p-1)!}\int d^{p}x\,\eps^{a_1\cdots a_{p}}\left(2F^{(p+1)}_{ia_2\cdots a_{p}y,a}\cR^a{}_{a_1}{}^{iy}- 2F^{(p+1)}_{aa_2\cdots a_{p}y,i}\cR^i{}_{a_1}{}^{ay}\right)\labell{first22}
\eeqa
where we have used the assumption in T-duality that all field are independent of the Killing direction $y$. The coordinate $y$ in the T-dual theory is  a transverse coordinate. Inspired by the above couplings, one may guess that the correct form of the couplings for D$_{p-1}$-brane are in fact,
\beqa
\frac{T_{p-1}}{(p-1)!}\int d^{p}x\,\eps^{a_1\cdots a_{p}}\left(F^{(p+1)}_{ia_2\cdots a_{p}j,a}\cR^a{}_{a_1}{}^{ij}- 2F^{(p+1)}_{aa_2\cdots a_{p}j,i}\cR^i{}_{a_1}{}^{aj}\right)\labell{first222}
\eeqa
This is consistent with the couplings \reef{second} that we have found from the S-matrix. 

The last term above is not  consistent with the T-duality along the world volume direction. To see this, consider the case that $a_1$ carries the index $y$. The world volume index $a$ should be separated into $y$ and $\ta$, which does not include the coordinate $y$. So the second term in \reef{first222} can be written as
\beqa
-\frac{2T_{p-1}}{(p-1)!}\int d^px\,\eps^{ya_2\cdots a_p}\left(F^{(p+1)}_{\ta a_2\cdots a_pj,i}\cR^i{}_y{}^{\ta j}+ F^{(p+1)}_{y a_2\cdots a_pj,i}\cR^i{}_y{}^{yj}\right)
\eeqa
Under the linear T-duality it transforms to
\beqa
-\frac{T_{p-1}}{(p-1)!}\int d^{p-1}x\,\eps^{a_2\cdots a_p}\left(F^{(p+2)}_{\ta a_2\cdots a_pjy,i}H^{\ta yj,i}-(-1)^pF^{(p)}_{a_2\cdots a_pj,i}h_{yy}{}^{,ij}\right)
\eeqa
The first term above, in particular,  indicates that there must be the following coupling:
\beqa
-\frac{T_{p}}{2!(p+1)!}\int d^{p+1}x\,\eps^{a_0\cdots a_p}F^{(p+4)}_{aa_0\cdots a_pjk,i}H^{akj,i}
\eeqa
However, this coupling is not produced by the S-matrix element \reef{K123}. To fix this inconsistency, we  use the on-shell conditions to rewrite the second term in \reef{first222} in a T-dual form. The index $a$ in this term can be only $a_1$  so we can rewrite it as $-2F^{(p+1)}_{a_1a_2\cdots a_{p}j,i}\cR^i{}_{a}{}^{aj}/p$. Moreover, using the on-shell conditions, one can write the curvature as $\hat{\cR}^{ij}$ where
\beqa
\hat{\cR}_{ij}&\equiv &\frac{1}{2}(\cR_{ia}{}^a{}_j-\cR_{ik}{}^k{}_j)
\eeqa
It does not have $a_1$ index anymore to produce inconsistency with T-duality. It has been shown in \cite{Garousi:2009dj} that it is invariant under linear T-duality transformations \reef{linear} when $i,j\ne y$.  Hence, the couplings \reef{first222} can be written for D$_p$-brane as
\beqa
2T_{p}\int d^{p+1}x\,\eps^{a_0\cdots a_{p}}\left(\frac{1}{2!p!}F^{(p+2)}_{ia_1\cdots a_pj,a}\cR^a{}_{a_0}{}^{ij}-\frac{1}{(p+1)!}F^{(p+2)}_{a_0\cdots a_pj,i}\hat{\cR}^{ij}\right)\labell{second1}
\eeqa
which are equivalent to the couplings \reef{second} using on-shell conditions. These are the couplings in the second line of \reef{LTdual}.

 It has been speculated in \cite{Garousi:2009dj} that the non-constant dilaton   appears  in the string frame D-brane action  in the same way that the constant dilaton appears in the action, \eg  the non-constant dilaton appears only through the overall factor of $e^{-\phi}$ in the string frame DBI action. This proposal has been verified for DBI action by explicit calculation at order $\alpha'^2$ in \cite{Garousi:2009dj}. We now check the proposal for the couplings that we have found. According to this proposal the dilaton appears only through the string frame metric in \reef{second1}.
 %In the CS part the speculation is that the dilaton appears only through the string frame metric.
% where in the Chern-Simons part, it is absorbed  in the RR field, \ie $e^{-\phi }C\rightarrow C$. 
In other words, the  dilaton couplings in the Einstein frame  should be given by  transforming the  string frame   couplings \reef{second1} to the Einstein frame, \ie replacing $h_{\mu\nu}\rightarrow \phi\eta_{\mu\nu}/2$. This replacement gives $\cR^a{}_{a_0}{}^{ij}\rightarrow 0$, $\cR^i{}_{a}{}^{aj}\rightarrow (\eta^{ij}\phi_{,a}{}^{a}+\eta_{a}{}^a\phi^{,ij})/4$ and $\cR^i{}_{k}{}^{kj}\rightarrow [\eta^{ij}\phi_{,k}{}^{k}+(\eta_{k}{}^k-2)\phi^{,ij}]/4$. Using the on-shell condition  that $F_{a_0\cdots a_pi}{}^{,i}=-F_{a_0\cdots a_pa}{}^{,a}=0$, one finds the following dilaton coupling:
\beqa
-\frac{T_p(p-3)}{2(p+1)!}\int d^{p+1}x\,\eps^{a_0\cdots a_p}F^{(p+2)}_{a_0\cdots a_pj,i}\phi\,^{,ij}
\eeqa
which is exactly the coupling \reef{dil}. Note that if one uses the replacement $h_{\mu\nu}\rightarrow \phi\eta_{\mu\nu}/2$ in the couplings \reef{second} which is consistent with S-matrix but not with T-duality, one would not find the correct dilaton coupling in the  Einstein frame. 

Having fixed the on-shell ambiguity of the last term in \reef{second} by requiring the consistency with linear T-duality, we now fix the on-shell ambiguity of the last term in \reef{first} by requiring that the  T-duality along a transverse direction of the equation \reef{second1} should produce the $F^{(p)}H$ couplings. Reversing the steps to find the first term in \reef{second1}, one finds the first term in \reef{first}. To find the T-dual of the last term in \reef{second1}, we write it as
\beqa
-\frac{2T_p}{(p+1)!}\int d^{p+1}x\,\eps^{a_0\cdots a_p}\left(F^{(p+2)}_{a_0\cdots a_p\tj,i}\tilde{\cR}^{i\tj}+ F^{(p+2)}_{a_0\cdots a_py,i}\tilde{\cR}^{iy}\right)
\eeqa
where the transverse index $\tj$ does not include $y$. The T-duality transformation of the first term above gives a term which is reproduced by the second term in \reef{second1}, and the T-duality of the second term gives the following terms:
\beqa
\frac{T_{p+1}}{2(p+1)!}\int d^{p+1}xdy\,\eps^{ya_0\cdots a_p}F^{(p+1)}_{a_0\cdots a_p,i}\left(H^{ia}{}_{y,a}-H^{ij}{}_{y,j}\right)
\eeqa
Inspired by these terms, one guesses that there must be the following couplings:
\beqa
\frac{T_{p}}{2p!}\int d^{p+1}x\,\eps^{a_1\cdots a_p}F^{(p+1)}_{a_1\cdots a_p,i}\left(H^{ia}{}_{a_0,a}-H^{ij}{}_{a_0,j}\right)
\eeqa
This fixes the on-shell ambiguity in the second term in \reef{first}. Hence, the couplings which are consistent with the S-matrix element and with the linear T-duality are those that appear in the first line of \reef{LTdual}.

Now we consider the transformation of the couplings \reef{second1} under linear T-duality transformations along a world volume direction. From the contraction with the world volume form, one of the indices  of the RR field strength or the index   of the curvature   in \reef{second1} must include $y$, and so again there are two cases to consider: First when $y$ appears as an index on the RR field strength and second when $y$ is an index in the curvature. In the former case, one can easily check that the T-dual couplings are consistent with \reef{second1}. In the latter case, we  write \reef{second1} as 
\beqa
2T_{p}\int d^{p+1}x\,\eps^{ya_1\cdots a_{p}}\left(\frac{1}{2!p!}F^{(p+2)}_{ia_1\cdots a_{p}j,a}\cR^a{}_{y}{}^{ij}\right)\labell{second2}
\eeqa
It  transforms under linear T-duality to the following coupling of D$_{p-1}$-brane:
\beqa
T_{p-1}\int d^{p}x\,\eps^{a_1\cdots a_{p}}\left(\frac{1}{2!p!}F^{(p+3)}_{ia_1\cdots a_{p}jy,a}H^{iyj,a}\right)\labell{second3}
\eeqa
Note that after T-duality the transverse indices $i,j$ should be written as $\ti,\tj$ which do not include $y$, however, because $H^{\ti y\ti}$ is totally antisymmetric we wrote it as $H^{iyj}$. Inspired by this couplings, one may guess that the correct form of the coupling for D$_{p}$-brane is in fact,
\beqa
T_{p}\int d^{p+1}x\,\eps^{a_0a_1\cdots a_{p}}\left(\frac{1}{3!(p+1)!}F^{(p+4)}_{ia_0\cdots a_{p}jk,a}H^{ikj,a}\right)\labell{second4}
\eeqa
which is  the coupling \reef{third} that we have found from the S-matrix element. This is the coupling in the last line of \reef{LTdual}. There is no index in the NSNS field strength in above coupling that contracts with the world volume form. So continuing  the T-duality along a world volume direction, one would find no new term involving $F^{p+6}$. This is consistent with the S-matrix calculation in the previous section that indicates there is no coupling for $F^{p+6}$. 
%Therefore, the couplings \reef{LTdual} are produce by disck level S-matrix elements and are consistent with the linear T-duality transformations.

\subsection{Non-linear T-duality }

We have seen in the previous section that the couplings \reef{LTdual} are consistent with the linear T-duality transformations \reef{linear}. However,  they are not consistent with nonlinear T-duality transformations. In this paper, we would like to study the effect of  the second   term  in the T-duality transformation of the RR potential in \reef{Cy}. So we consider the following T-duality transformations:
\beqa
&& \tilde{h}_{\mu y}=B_{\mu y},\, \tilde{B}_{\mu y}=h_{\mu y},\,\tilde{h}_{\mu\nu}=h_{\mu\nu},\,\tilde{B}_{\mu\nu}=B_{\mu\nu},\,\tC^{(n)}_{\mu\cdots\nu\alpha\beta}=nC^{(n-1)}_{[\mu\cdots\nu\alpha}B^{}_{\beta]y}\labell{nonlinear}
\eeqa
The reason for choosing only the nonlinear term above is that the consistency of the RR potential $C$ with this term makes it to be $e^BC$. To see this, consider   the linear coupling of the CS action \reef{CS2}, \ie
\beqa
T_p\int C^{(p+1)}&=&\frac{T_p}{(p+1)!}\int d^{p+1}x\,\eps^{a_0\cdots a_p}C^{(p+1)}_{a_0\cdots a_p}
\eeqa
The transformation of the above coupling of D$_p$-brane under T-duality along a transverse direction  gives the following coupling for D$_{p+1}$-brane:
\beqa
\frac{T_{p+1}}{p!}\int d^{p+1}xdy\,\eps^{a_0\cdots a_py}C^{(p)}_{a_0\cdots a_{p-1}}B_{a_{p}y}
\eeqa
This dictates that there must be the following coupling:
\beqa
\frac{T_{p}}{2!(p-1)!}\int d^{p+1}x\,\eps^{a_0\cdots a_p}C^{(p-1)}_{a_0\cdots a_{p-2}}B_{a_{p-1}a_p}&=&T_p\int C^{(p-1)}\wedge B
\eeqa
which is a standard term in the CS action \reef{CS2}.

We now study the consistency of the couplings in the first line of \reef{LTdual} under the T-duality transformation \reef{nonlinear}.  To study the effect of last term in \reef{nonlinear}, we have to consider the couplings in which the RR field carries no  index $y$.  We begin by implementing T-duality along a transverse direction of the D$_p$-brane. The B-fields in the first line of \reef{LTdual} are invariant under \reef{nonlinear}, so 
these couplings  transform under the T-duality to the following couplings for D$_{p+1}$-brane:
\beqa
\frac{T_{p+1}}{2!(p-1)!}\int d^{p+1}xdy\,\eps^{a_0a_1\cdots a_{p}y}\left(\tilde{F}^{(p)}_{ia_2\cdots a_{p},a}{H}_{a_0a_1}{}^{a,i}- \frac{1}{p}\tilde{F}^{(p)}_{a_1\cdots a_p,i}(H_{a_0a}{}^{i,a}-H_{a_0j}{}^{i,j})\right)\labell{first11}
\eeqa
Let us first consider the second term above. The transformation of the RR field strength $\tilde{F}^{(p)}_{a_1\cdots a_{p}}$ under the T-duality \reef{nonlinear} is
\beqa
\tilde{F}^{(p)}_{a_1\cdots a_{p}}&=&p(\tilde{C}^{(p-1)}_{a_2\cdots a_{p}})_{,a_1}\nonumber\\
%&=&p(p-1)(C^{(p-2)}_{a_2\cdots a_{p-1}}B_{a_py})_{,a_1}\nonumber\\
&=&pF^{(p-1)}_{a_1\cdots a_{p-1}}B_{a_py}+\frac{p(p-1)}{2}C^{(p-2)}_{a_2\cdots a_{p-1}}H_{a_pya_1}\labell{tF}
\eeqa
where here and in the subsequent identities we have used the fact that the world volume indices $a_0,a_1,\cdots a_p$  are contracted with the totally antisymmetric  world volume tensor. Inserting this in the second term in \reef{first11}, one finds that there must be the following couplings:
\beqa
&&-\frac{T_{p+1}}{2!}\int d^{p+1}xdy\,\eps^{a_0a_1\cdots a_{p+1}}\left(\frac{1}{2!(p-1)!}{F}^{(p-1)}_{a_1\cdots a_{p-1}}B_{a_pa_{p+1}}+ \frac{1}{3!(p-2)!}{C}^{(p-2)}_{a_2\cdots a_{p-1}}H_{a_pa_{p+1}a_1}\right)_{,i}\nonumber\\
&&\qquad\qquad\qquad\qquad\qquad\qquad\times (H_{a_0a}{}^{i,a}-H_{a_0j}{}^{i,j})\labell{first110}
\eeqa
The terms in the bracket in the first line is
\beqa
\frac{1}{(p+1)!}(B\wedge F^{(p-1)})_{a_1\cdots a_{p+1}}+\frac{1}{(p+1)!}(H\wedge C^{(p-2)})_{a_1\cdots a_{p+1}}
\eeqa
Hence, the couplings \reef{first110} are given by the second term in \reef{LTdual} in which the RR field strength is \reef{newF}.

The transformation of the RR field strength $\tilde{F}^{(p)}_{ia_2\cdots a_{p}}$ in the first term of \reef{first11} is
\beqa
\tilde{F}^{(p)}_{ia_2\cdots a_{p}}&=&\tC^{(p-1)}_{a_2\cdots a_p,i}-(p-1)\tC^{(p-1)}_{ia_3\cdots a_p,a_2}\nonumber\\
%&=&(p-1)(C^{(p-2)}_{[a_2\cdots a_{p-1}}B^{}_{a_p]y})_{,i}-(p-1)^2(C^{(p-2)}_{[ia_3\cdots a_{p-1}}B^{}_{a_p]y})_{,a_2}\nonumber\\
%&=&(p-1)[(C^{(p-2)}_{a_2\cdots a_{p-1}}B_{a_py})_{,i}-(p-2)(C^{(p-2)}_{ia_3\cdots a_{p-1}}B_{a_py})_{,a_2}+(-1)^{p-2}(C^{(p-2)}_{a_3\cdots a_p}B_{iy})_{,a_2}]\nonumber\\
&=&(p-1)F^{(p-1)}_{ia_2\cdots a_{p-1}}B_{a_py}+(-1)^{p-2}F^{(p-1)}_{a_2\cdots a_p}B_{iy}\labell{tF2}\\
&&+(-1)^{p-2}\frac{(p-1)(p-2)}{2}C^{(p-2)}_{ia_2\cdots a_{p-2}}H_{a_{p-1}a_py}+(p-1)C^{(p-2)}_{a_2\cdots a_{p-1}}H_{a_pyi}\nonumber
\eeqa
Inserting this in equation \reef{first11}, one finds that there must be the following new terms:
\beqa
&&\frac{T_{p+1}}{2!}\int d^{p+1}xdy\,\eps^{a_0a_1\cdots a_{p}a_{p+1}}\,H_{a_0a_1}{}^{a,i}\left(\frac{1}{2!(p-2)!}F^{(p-1)}_{ia_2\cdots a_{p-1}}B_{a_pa_{p+1}}+\frac{1}{(p-1)!}F^{(p-1)}_{a_3\cdots a_{p+1}}B_{ia_2}\right.\nonumber\\
&&\left.-\frac{1}{3!(p-3)!}C^{(p-2)}_{a_2\cdots a_{p-2}i}H_{a_{p-1}a_pa_{p+1}}+\frac{1}{2!(p-2)!}C^{(p-2)}_{a_2\cdots a_{p-1}}H_{a_pa_{p+1}i}\right)_{,a}\labell{first111}
\eeqa
%The new terms inspired by the second term in \reef{first111} are the same as above in which $i\leftrightarrow a$. 
Consider the following identities:
\beqa
\frac{1}{p!}(B\wedge F^{(p-1)})_{ia_2\cdots a_{p+1}}&=&\frac{1}{(p-1)!}B_{ia_2}F^{(p-1)}_{a_3\cdots a_{p+1}}-\frac{1}{2!(p-2)!}B_{a_3a_2}F^{(p-2)}_{ia_4\cdots a_{p+1}}\nonumber\\
\frac{1}{p!}(H\wedge C^{(p-2)})_{ia_2\cdots a_{p+1}}&=&\frac{1}{2!(p-1)!}H_{ia_2a_3}C^{(p-2)}_{a_4\cdots a_{p+1}}-\frac{1}{3!(p-3)!}H_{a_2a_3a_4}C^{(p-2)}_{ia_5\cdots a_{p+1}}
\eeqa
The sum of the above terms gives exactly the terms in the bracket in \reef{first111}. Hence the new terms \reef{first111} are given by  the coupling in  the first term of \reef{LTdual} in which the RR field strength is given by \reef{newF}.

Now consider T-duality of the couplings in the first line of \reef{LTdual} along a world volume  direction. From the contraction with the world volume form, one of the indices  $a_0,a_1$ of the NSNS field strength   must include $y$. Note that the RR field has no $y$ index in the nonlinear T-duality transformation \reef{nonlinear}. The T-duality on the  B-field is the same as in the previous section, so the couplings  transform under the T-duality \reef{nonlinear} to the following couplings for D$_{p-1}$-brane:
\beqa
2\frac{T_{p-1}}{(p-1)!}\int d^{p}x\,\eps^{a_1\cdots a_{p}}\left(\tilde{F}^{(p)}_{ia_2\cdots a_{p},a}\cR^a{}_{a_1}{}^{iy}- \frac{1}{p}\tilde{F}^{(p)}_{a_1\cdots a_{p},i}\hat{\cR}^{iy}\right)\labell{nfirst22}
\eeqa
The transformation of the RR field strengths are  given by \reef{tF2} and \reef{tF}.  However, the $y$ coordinate is now a transverse coordinate, so unlike the previous case the new terms inspired by the above couplings can not be written as $B\wedge F+H\wedge C$. This indicates that there must be some other contributions as well. The other contributions are coming from the T-duality transformation of the couplings in the second line of \reef{LTdual} along a transverse direction. The transformation of these terms under \reef{nonlinear} gives the following couplings for D$_{p+1}$-brane:
\beqa
2\frac{T_{p+1}}{p!}\int d^{p}xdy\,\eps^{a_0\cdots a_{p}y}\left(\frac{1}{2!}\tilde{F}^{(p+2)}_{ia_1\cdots a_{p}j,a}\cR^a{}_{a_0}{}^{ij}- \frac{1}{p+1}\tilde{F}^{(p+2)}_{a_0\cdots a_{p}j,i}\hat{\cR}^{ij}\right)\labell{nfirst222}
\eeqa
The new couplings inspired by the couplings \reef{nfirst22} and \reef{nfirst222} should be given by the couplings in the second line of \reef{LTdual} in which the RR field strength is \reef{newF}.

Similarly, the T-duality of the couplings in the second line of \reef{LTdual} along a world volume direction is given by the following coupling for D$_{p-1}$-brane:
\beqa
T_{p-1}\int d^{p}x\,\eps^{a_1\cdots a_{p}}\left(\frac{1}{2!p!}\tilde{F}^{(p+2)}_{ia_1\cdots a_pj,a}H^{iyj,a}\right)\labell{second11}
\eeqa
and the T-duality of the coupling in the last line of \reef{LTdual} along a transverse direction is given by the following coupling for D$_{p+1}$-brane:
\beqa
T_{p+1}\int d^{p+1}xdy\,\eps^{a_0\cdots a_{p}y}\left(\frac{1}{3!(p+1)!}\tilde{F}^{(p+4)}_{i a_0\cdots a_{p}jk,a}H^{ikj,a}\right)\labell{second42}
\eeqa
The new couplings inspired by the couplings \reef{second11} and \reef{second42} should be given by the couplings in the last line of \reef{LTdual} in which the RR field strength is \reef{newF}. Let us check this case explicitly.

%Now consider the couplings in \reef{second1}. The linear curvature term with no index along the Killing direction is  invariant under the linear T-duality transformation. The graviton however is not invariant under the  nonlinear T-duality transformation \reef{h}. This  indicates that the linear curvature is not invariant under the nonlinear T-duality transformation. However, we expect the nonlinear curvature term with no index along the Killing direction to be invariant under the nonlinear T-duality transformation, \eg the nonlinear T-duality transformation of  the linear terms of the curvature should be canceled with the linear T-duality transformation of the nonlinear terms of the curvature. Using this assumption, one observes that implementing T-duality of the couplings \reef{second1} along a transverse direction gives no coupling  of one RR and two B-fields in which we are interested. So we implement the T-duality along a world volume direction. From the contraction with the world volume form, one of the indices  of the RR field strength or the index   of the curvature in \reef{second1} must include $y$. Only the latter case can produce  the couplings of  one RR and two B-fields under T-duality. This part of the T-duality transformation of \reef{second1} is then given by the following couplings for D$_{p-1}$-brane: 
 Shifting $p\rightarrow p+1$ in \reef{second11}, one can write this equation as
\beqa
T_{p}\int d^{p+1}x\,\eps^{a_0\cdots a_{p}}\left(\frac{1}{2!(p+1)!}\tilde{F}^{(p+3)}_{ia_0\cdots a_pj,a}H^{iyj,a}\right)\labell{second111}
\eeqa
The transformation of the RR field strength $\tilde{F}^{(p+3)}_{ia_0\cdots a_pj}$ under \reef{nonlinear} is
\beqa
\tilde{F}^{(p+3)}_{ia_0\cdots a_pj}&=&2\tC^{(p+2)}_{a_0\cdots a_pj,i}-(p+1)\tC^{(p+2)}_{ia_1\cdots a_pj,a_0}\nonumber\\
%&=&2(p+2)(C^{(p+1)}_{[a_0\cdots a_p}B^{}_{j]y})_{,i}-(p+1)(p+2)(C^{(p+1)}_{[ia_1\cdots a_p}B^{}_{j]y})_{,a_0}\nonumber\\
&=&2F^{(p+2)}_{ia_0\cdots a_p}B_{jy}-(p+1)F^{(p+2)}_{ia_0\cdots a_{p-1}j}B_{a_py}\nonumber\\
&&-2(p+1)C^{(p+1)}_{a_0\cdots a_{p-1}j}H_{a_pyi}+C^{(p+1)}_{a_0\cdots a_p}H_{jyi}+\frac{p(p+1)}{2}C^{(p+1)}_{a_0\cdots a_{p-2}ij}H_{a_pya_{p-1}}
\eeqa
Inserting this in equation \reef{second111}, one finds that there should be the following new terms:
\beqa
&&T_{p}\int d^{p+1}x\,\eps^{a_0\cdots a_{p}}\left([\frac{1}{2!(p+1)!}F^{(p+2)}_{ia_0\cdots a_p}B_{jk}-\frac{1}{2!p!}F^{(p+2)}_{ia_0\cdots a_{p-1}j}B_{a_pk}\right.\labell{one}\\
&&\left.-\frac{1}{2!p!}C^{(p+1)}_{a_0\cdots a_{p-1}j}H_{a_pki}+\frac{1}{3!(p+1)!}C^{(p+1)}_{a_0\cdots a_p}H_{jki}+\frac{1}{2!2!(p-1)!}C^{(p+1)}_{a_0\cdots a_{p-2}ij}H_{a_{p-1}a_pk}]_{,a}H^{ikj,a}\right)\nonumber
\eeqa
Checking the indices, one realizes that the above terms are not given by the last term of \reef{LTdual} in which $F$ is replaced by \reef{newF}. Now  shifting $p\rightarrow p-1$ in equation \reef{second42}, one can write it as
\beqa
T_{p}\int d^{p}xdy\,\eps^{a_0\cdots a_{p-1}y}\left(\frac{1}{3!(p)!}\tilde{F}^{(p+3)}_{i a_0\cdots a_{p-1}jk,a}H^{ikj,a}\right)\labell{second421}
\eeqa
The transformation of the RR field strength $\tilde{F}^{(p+3)}_{i a_0\cdots a_{p-1}jk}$ under \reef{nonlinear} is
\beqa
\tilde{F}^{(p+3)}_{i a_0\cdots a_{p-1}jk}&=&3\tC^{(p+2)}_{a_0\cdots a_{p-1}jk,i}-p\tC^{(p+2)}_{ia_1\cdots a_{p-1}jk,a_0}\nonumber\\
%&=&3(p+2)(C^{(p+1)}_{[a_0\cdots a_{p-1}j}B^{}_{k]y})_{,i}-p(p+2)(C^{(p+1)}_{[ia_1\cdots a_{p-1}j}B^{}_{k]y})_{,a_0}\nonumber\\
&=&3F^{(p+2)}_{ia_0\cdots a_{p-1}j}B_{ky}+pF^{(p+2)}_{ia_0\cdots a_{p-2}jk}B_{a_{p-1}y}\nonumber\\
&&+3pC^{(p+1)}_{a_0\cdots a_{p-2}jk}H_{a_{p-1}yi}+3C^{(p+1)}_{a_0\cdots a_{p-1}j}H_{kyi}+\frac{p(p-1)}{2}C^{(p+1)}_{ia_1\cdots a_{p-3}a_0jk}H_{a_{p-2}a_{p-1}y}\nonumber
\eeqa
Inserting this in equation \reef{second421}, one finds that the first , third and fourth terms above are reproduced by the new couplings  \reef{one} when one chooses $a_p=y$ in \reef{one}.  The other two terms led us to guess that there should be the following new terms:
\beqa
&&T_{p}\int d^{p+1}x\,\eps^{a_0\cdots a_{p}}\left([\frac{1}{2!3!(p-1)!}F^{(p+2)}_{ia_0\cdots a_{p-2}jk}B_{a_{p-1}a_p}\right.\labell{two}\\
%&&\left.\qquad\qquad\qquad\qquad+\frac{1}{2!2!(p-1)!}C_{a_0\cdots a_{p-2}jk}H_{a_{p-1}a_pi}+\frac{1}{2!p!}C_{a_0\cdots a_{p-1}j}H_{ka_pi}\right.\nonumber\\
&&\left.\qquad\qquad\qquad\qquad-\frac{1}{3!3!(p-2)!}C^{(p+1)}_{a_0\cdots a_{p-3}ijk}H_{a_{p-2}a_{p-1}a_p}]_{,a}H^{ikj,a}\right)\nonumber
\eeqa
%where the terms that have $B_{a_0a_p},\cdots B_{a_{p-1}a_p}$ one must divide them by 2.
One can easily check that the new couplings \reef{one} and \reef{two} are reproduced  by the last term in \reef{LTdual} in which the RR field strength is given by $B\wedge F^{(p+2)}+H\wedge C^{(p+1)}$.

Now consider the new couplings in \reef{LTdual} which have one RR and two NSNS states. If one implements the T-duality on these terms and use the nonlinear T-duality transformations \reef{nonlinear}, one should find new terms which are given by \reef{LTdual} in which the RR field strength is given by the terms in the second line of \reef{newF}. Let us check this for the second term in \reef{LTdual} which is simple to analyze. Implementing the T-duality along a transverse direction, one finds the following couplings for D$_{p+1}$-brane:
\beqa
&&-\frac{T_{p+1}}{2}\int d^{p+1}xdy\,\eps^{a_0\cdots a_py}(H_{a_0a}{}^{i,a}-H_{a_0j}{}^{i,j})\labell{final}\\
&&\times\left(\frac{1}{2!(p-2)!}B_{a_1a_2}\tilde{F}^{(p-2)} _{a_3\cdots a_p}+\frac{1}{3!(p-4)!}H_{a_1a_2a_3}C^{(p-4)}_{a_4\cdots a_{p-1}}B_{a_py}\right)_{,i}\nonumber
\eeqa  
where the T-duality of the RR field strength is given in \reef{tF}. Inspired by the above equation, one guesses that there must be the following couplings:
\beqa
&&-\frac{T_{p+1}}{2}\int d^{p+1}xdy\,\eps^{a_0\cdots a_py}(H_{a_0a}{}^{i,a}-H_{a_0j}{}^{i,j})\left(\frac{1}{2!2!2!(p-3)!}B_{a_1a_2}F^{(p-3)}_{a_3\cdots a_{p-1}}B_{a_pa_{p+1}}\right.\nonumber\\
&&\left.+\frac{1}{2!2!3!(p-4)!}B_{a_1a_2}C^{(p-4)}_{a_4\cdots a_{p-1}}H_{a_pa_{p+1}a_3}+\frac{1}{2!2!3!(p-4)!}H_{a_1a_2a_3}C^{(p-4)}_{a_4\cdots a_{p-1}}B_{a_pa_{p+1}}\right)_{,i}\nonumber
\eeqa
Shifting $p+1\rightarrow p$, one finds that the above couplings are exactly given by the second term in \reef{LTdual} in which the RR field strength is given by the second line of \reef{newF}. This ends our illustration of the consistency between the couplings \reef{LTdual} and the T-duality.

We have seen that the consistency with a particular term of the  nonlinear T-duality  guides  us to write the RR field strength in \reef{LTdual} as ${\cal F}=d{\cal C}$. The resulting couplings however are consistent with all nonlinear terms of the RR potential. In fact the nonlinear T-duality transformation \reef{Cy} for the RR potential can be written as \cite{Taylor:1999pr}
\beqa
 \tilde{\cal C}^{(n)}_{\mu\cdots \nu\alpha y}={\cal C}^{(n-1)}_{\mu\cdots \nu\alpha }&,&\tilde{\cal C}^{(n)}_{\mu\cdots\nu\alpha\beta}={\cal C}^{(n+1)}_{\mu\cdots\nu\alpha\beta y}\labell{nonlinear2}
\eeqa
The calculations in section 3.1 then show that the couplings \reef{LTdual} with the RR field strength ${\cal F}=d{\cal C}$ are consistent with the full nonlinear T-duality transformation for the RR field. It would be interesting to confirm  the couplings of one RR and two NSNS fields in \reef{LTdual} by the disk level scattering amplitude.
%\section{Discussion}

{\bf Acknowledgments}: I would like to thank Katrin Becker and Rob Myers for  useful  discussions. 
 This work is supported by Ferdowsi University of Mashhad under grant p/964(1388/12/4). 
%%%%%%%%%%%%%%%%%%%%%%%%%%%%%%%%%%%%
%%%%%%%%%%%%%%%%%%%%%%%%%%%%%%%%%%%%
\bibliographystyle{/Users/Nick/utphys} %\bibliography{/Users/Nick/myrefs}
\bibliographystyle{utphys} \bibliography{hyperrefs-final}
%%%%%%%%%%%%%%%%%%%%%%%%%%%%%%%%%%%%

%%%%%%%%%%%%%%%%%%%%%%%%%%%%%%%%%%%%
%%%%%%%%%%%%%%%%%%%%%%%%%%%%%%%%%%%%
%\bibliographystyle{/Users/Nick/utphys} %\bibliography{/Users/Nick/myrefs}
%\bibliographystyle{utphys} \bibliography{hyperrefs-final}
%%%%%%%%%%%%%%%%%%%%%%%%%%%%%%%%%%%%

\providecommand{\href}[2]{#2}\begingroup\raggedright

%\newpage
\endgroup

\end{document}